\documentclass[12pt,twocolumn,Letter]{iopart}
\usepackage{iopams}
\usepackage{graphicx}
\usepackage{mathtext}

\begin{document}

\newcommand {\eps} {\varepsilon}
\newcommand {\wD} {\widetilde{D}}
\newcommand {\ws} {\widetilde{\sigma}}
\newcommand {\wq} {\widetilde{q}}
\newcommand {\wu} {\widetilde{u}}
\newcommand {\la} {\left\langle}
\newcommand {\ra} {\right\rangle}

\title{Advectional enhancement of eddy diffusivity under parametric disorder}

\author{Denis S Goldobin}
\address{Department of Theoretical Physics, Perm State University,
        15 Bukireva str., 614990, Perm, Russia}
\ead{Denis.Goldobin@gmail.com}

\begin{abstract}
Frozen parametric disorder can lead to appearance of sets of
localized convective currents in an otherwise stable (quiescent)
fluid layer heated from below. These currents significantly
influence the transport of an admixture (or any other passive
scalar) along the layer. When the molecular diffusivity of the
admixture is small in comparison to the thermal one, which is
quite typical in nature, disorder can enhance the effective (eddy)
diffusivity by several orders of magnitude in comparison to the
molecular diffusivity. In this paper we study the effect of an
imposed longitudinal advection on delocalization of convective
currents, both numerically and analytically; and report subsequent
drastic boost of the effective diffusivity for weak advection.
\end{abstract}

\pacs{05.40.-a,    
      44.30.+v,    
      47.54.-r,    
      72.15.Rn     
}
\vspace{2pc}
\noindent{\it Special Issue}: Article preparation, IOP journals\\
\submitto{\it \PS}
\maketitle


Disorder in operation conditions of a dynamic system is known to
be able to play not only trivial destructive role, distorting the
system behavior, but also constructive one, inducing certain
degree of order and leading to various non-trivial effects:
Anderson
localization~\cite{Anderson-1958,Froehlich-Spencer-1984,Rossum-Nieuwenhuizen-1999,Maynard-2001},
stochastic~\cite{StochRes} and coherence resonances~\cite{CohRes},
noise-induced synchronization~\cite{NoiseSynch}, etc. One of the
most distinguished and fair effects is the Anderson localization
(AL), which is the localization of states in spatially extended
linear systems subject to a frozen parametric disorder (random
spacial inhomogeneity of parameters). AL was first discovered and
discussed for quantum systems~\cite{Anderson-1958}. Later on,
investigations were extended to diverse branches of semiclassical
and classical physics: wave
optics~\cite{Rossum-Nieuwenhuizen-1999},
acoustics~\cite{Maynard-2001}, etc. The phenomenon was
comprehensively studied and well understood mathematically for the
Schr\"odinger equation and related mathematical models
\cite{Froehlich-Spencer-1984,Lifshitz-Gredeskul-Pastur-1988,Gredeskul-Kivshar-1992}.
The role of nonlinearity in these models was addressed in
literature as well (e.g.,
\cite{Gredeskul-Kivshar-1992,Pikovsky-Shepelyansky-2008}).

While extensively studied for conservative media, the localization
phenomenon did not receive comparable attention for
active/dissipative ones, like in problems of thermal convection or
reaction-diffusion. The main reason is that the physical
interpretation of formal solutions to the Schr\"odinger equation
is essentially different from that of governing equations for
active/dissipative media; therefore, the theory of AL may be
extended to the latter only under certain strong restrictions
(see~\cite{Goldobin-Shklyaeva-PRE,Goldobin-Shklyaeva-2009} for
reference). Nevertheless, effects similar to AL can be observed in
fluid dynamical
systems~\cite{Goldobin-Shklyaeva-PRE,Goldobin-Shklyaeva-2009}.
In~\cite{Goldobin-Shklyaeva-2009} we addressed the problem where
localized thermoconvective currents excited in a horizontal porous
layer under frozen parametric disorder (spacial inhomogeneity of
the macroscopic permeability, the heat diffusivity, etc.)
drastically influence the process of transport of a passive scalar
(e.g., a pollutant) along the layer. Below the threshold of
instability of the disorder-free system, the effective diffusivity
quantifying this transport has been found to be faithfully
determined by localization properties of patterns. Meanwhile,
in~\cite{Goldobin-Shklyaeva-PRE} these properties have been
revealed to be greatly affected by a weak imposed longitudinal
advection. Hence, one can expect a weak advection to lead to a
significant enhancement of the effective diffusivity of a nearly
indiffusive pollutant. Treatment of this effect is the subject of
the present paper.

The paper is organized as follows. In \sref{sec1} we formulate the
specific physical problem that we deal with and introduce the
relevant mathematical model; in particular, we discuss
disorder-induced excitation of localized currents below the
instability threshold of the disorder-free system and advectional
delocalization of these patterns. \Sref{sec2} presents the results
of a numerical simulation and calculation of the effective
diffusivity. In
\sref{sec3} we develop an analytical theory for the effective
diffusivity in the presence of an imposed advection. \Sref{concl}
ends the paper with conclusions.

\section{Problem formulation and current state of research}\label{sec1}
In nature and technology, a broad variety of active media where
pattern selection occurs is governed by Kuramoto--Sivashinsky type
equations. In the presence of an imposed advectional transport $u$
in the $x$-direction the modified Kuramoto--Sivashinsky equation
reads
\begin{equation}
\dot{\theta}(x,t)=-\left(u\,\theta(x,t)+\theta_{xxx}(x,t)
 +q(x)\,\theta_x(x,t)-(\theta_x(x,t))^3\right)_x.
\label{eq1-01}
\end{equation}
\noindent
This equation describes two-dimensional large-scale natural
thermal convection in a horizontal fluid layer heated from
below~\cite{Knobloch-1990,Shtilman-Sivashinsky-1991} and is still
valid for a turbulent fluid~\cite{Aristov-Frick-1989}, a binary
mixture at small Lewis number~\cite{Schoepf-Zimmermann-1989-1993},
a porous layer saturated with a
fluid~\cite{Goldobin-Shklyaeva-BR-2008}, etc. In these fluid
dynamical systems, except for the turbulent
one~\cite{Aristov-Frick-1989}, the plates bounding the layer
should be nearly thermally insulating (in comparison to the fluid)
for a large-scale convection to arise. In the problems mentioned,
equation~\eref{eq1-01} governs evolution of temperature
perturbations $\theta$ which are nearly uniform along the vertical
coordinate $z$ and determine fluid currents.

The origin of such a frequent occurrence of equation~\eref{eq1-01}
is its general validity, which may be argued as follows. Basic
laws in physics are conservation ones. This often results in final
governing equations having the form
$\partial_t[\mbox{quantity}]+\nabla\!\cdot\![\mbox{flux of
quantity}]=0$. Such conservation laws lead to
Kuramoto--Sivashinsky type equations. While the original
Kuramoto--Sivashinsky equation has a quadratic nonlinear term
(cf~\cite{Michelson-1986}), this term should be replaced by a
cubic one for the systems with the sign inversion symmetry of the
fields, which is widespread in nature, or for description of a
spatiotemporal modulation of an oscillatory mode. Thus the
governing equation takes the form~\eref{eq1-01}. On these grounds,
we state that equation~\eref{eq1-01} describes pattern formation
in a broad variety of physical systems.

We restrict this paper to the case of convection in a porous
medium~\cite{Goldobin-Shklyaeva-BR-2008,Goldobin-Shklyaeva-2009}
for the sake of definiteness; nonetheless, most of our results may
be extended in a straightforward manner to the other physical
systems mentioned.
\Eref{eq1-01}
is already dimensionless and below we introduce all parameters and
variables in appropriate dimensionless forms.

\begin{figure}[!t]
\center{
\begin{tabular}{cc}
  \sf (a)&\hspace{-10mm}\includegraphics[width=0.85\textwidth]%
 {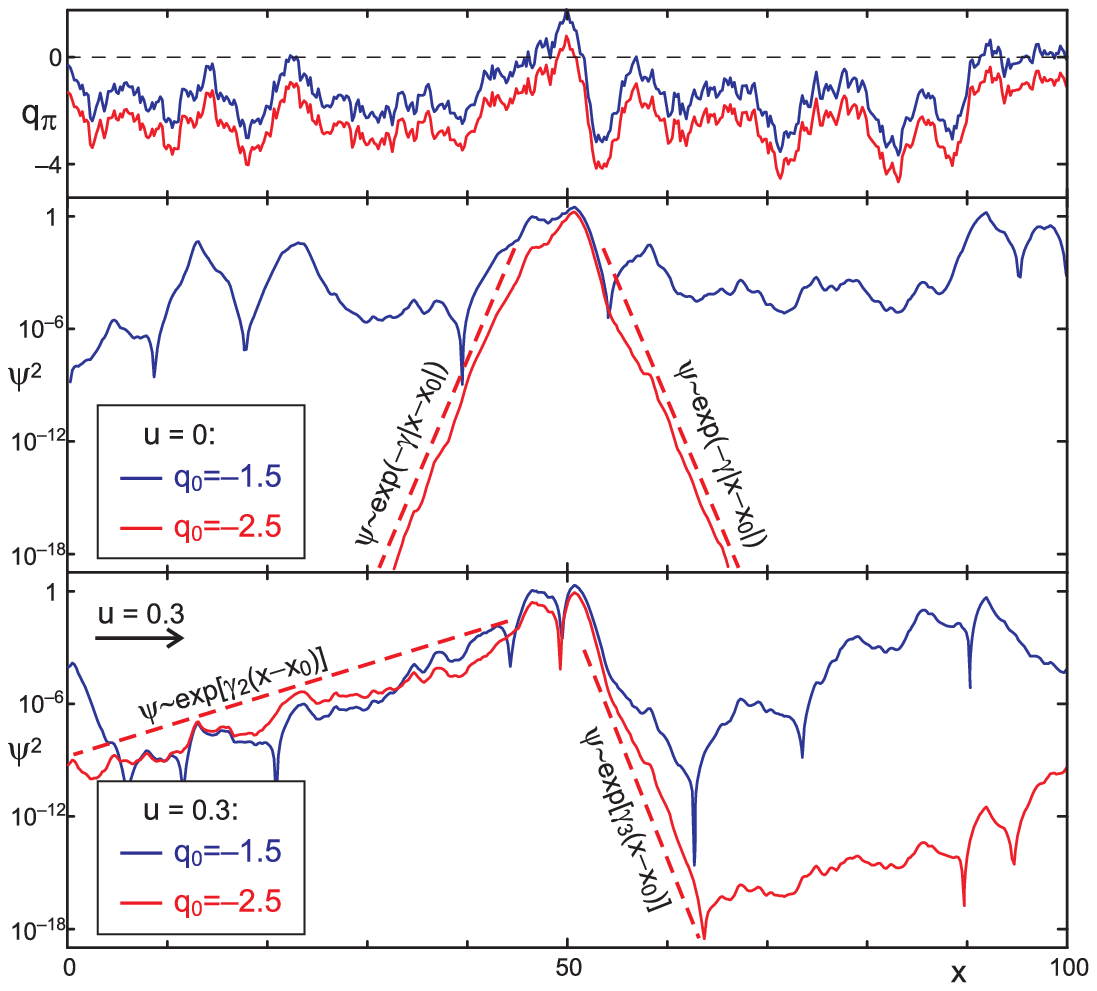}\\[5pt]
  \sf (b)&\hspace{-10mm}\includegraphics[width=0.85\textwidth]%
 {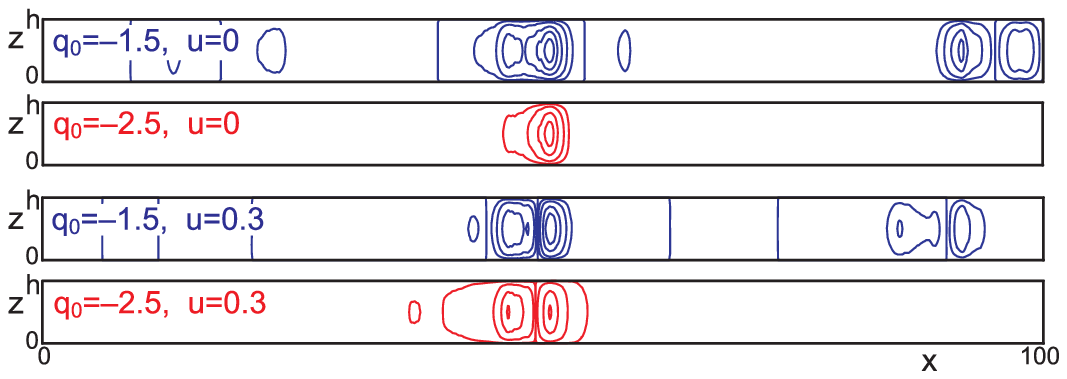}
\end{tabular}
}
  \caption{
(a) Establishing steady solutions to equation~\eref{eq1-01} are
sets of exponentially localized patterns [shown for one and the
same realization of random inhomogeneity $\xi(x)$ and $\eps=1$;
$q(x)$ is represented by
$q_\pi(x)=\pi^{-1}\int_{x-\pi/2}^{x+\pi/2}q(x')\rmd x'$].
\\
(b) The stream lines corresponding to the solutions in graph~(a)
are plotted for the case of convection in a porous layer [see
equation~\eref{eq1-02} for an exact relation].}
  \label{fig1}
\end{figure}

In the large-scale (or long-wavelength) approximation, which we
use, the characteristic horizontal scales are assumed to be large
against the layer height $h$.  In equation \eref{eq1-01}, $q(x)$
represents the local supercriticality: $(21/2)h^2q(x)$ is the sum
of relative deviations of the heating intensity and of the
macroscopic properties of the porous matrix (porosity,
permeability, heat diffusivity, etc.) from the critical values for
the spatially homogeneous case~\cite{Goldobin-Shklyaeva-BR-2008}.
For positive spatially uniform $q$, convection sets up, while for
negative $q$, all the temperature perturbations decay. In a porous
medium~\cite{Goldobin-Shklyaeva-BR-2008}, the macroscopic fluid
velocity field is
\begin{equation}
\vec{v}=\frac{\partial\Psi}{\partial z}\vec{e}_x
 -\frac{\partial\Psi}{\partial x}\vec{e}_z\,,
 \qquad
\Psi=\frac{3\sqrt{35}}{h^3}\,z(h-z)\,\theta_x(x,t)\equiv f(z)\,\psi(x,t)\,,
\label{eq1-02}
\end{equation}
\noindent
where $\psi(x,t)\equiv\theta_x(x,t)$ is the stream function
amplitude, and the reference frame is such that $z=0$ and $z=h$
are the lower and upper boundaries of the layer, respectively
[\fref{fig1}(b)]. Though the temperature perturbations obey
equation~\eref{eq1-01} for diverse convective systems, function
$f(z)$, which determines the relation between the flow pattern and
the temperature perturbation, is specific to each case. Notice
that $u$ is not presented in expression~\eref{eq1-02} owing to its
smallness in comparison to the excited convective currents
$\vec{v}$. The impact of a weak imposed advective flow on the
evolution of temperature perturbations is caused by its symmetry
properties: the gross advective flux through the vertical
cross-section is $u$, while the convective flow $\vec{v}$
possesses zero gross flux and, therefore, yields a less effective
heat transfer along the layer~\cite{Goldobin-Shklyaeva-BR-2008}.

Although equation \eref{eq1-01} is valid for a large-scale
inhomogeneity $q(x)$, which means $h|q_x|/|q|\ll1$, one can set
such a hierarchy of small parameters, namely
$h\ll(h|q_x|/|q|)^2\ll1$, that a frozen random inhomogeneity may
be represented by white Gaussian noise $\xi(x)$:
\[
q(x)=q_0+\xi(x),\quad \la\xi(x)\ra=0,\quad
\la\xi(x)\xi(x')\ra=2\eps^2\delta(x-x'),
\]
\noindent
where $\eps^2$ is the disorder intensity and $q_0$ is the mean
supercriticality (i.e.\ departure from the instability threshold
of the disorder-free system). Numerical simulation reveals only
time-independent solutions to establish in \eref{eq1-01} with
$u=0$ and such $q(x)$~\cite{Goldobin-Shklyaeva-PRE}; for a small
non-zero $u$, stable oscillatory regimes are of low probability by
continuity.

In the stationary case for $u=0$ the linearized form of equation
\eref{eq1-01}, i.e.,
\begin{equation}
-\theta_{xxx}(x)-\xi(x)\,\theta_x(x)=q_0\,\theta_x(x)\,,
\label{eq1-03}
\end{equation}
is a stationary Schr\"odinger equation for $\psi=\theta_x$ with
$q_0$ instead of the state energy and $-\xi(x)$ instead of the
potential. Therefore, similarly to the case of the Schr\"odinger
equation
(see~\cite{Froehlich-Spencer-1984,Lifshitz-Gredeskul-Pastur-1988,Gredeskul-Kivshar-1992}),
all the solutions $\psi(x)$ to the stationary linearized equation
\eref{eq1-01} are spatially localized for arbitrary $q_0$;
asymptotically,
\[
\psi(x)\propto\exp(-\gamma|x|),
\]
where $\gamma$ is the localization exponent. Such a localization
can be readily seen for the solution to the nonlinear problem
\eref{eq1-01} in \fref{fig1}(a) for $q_0=-2.5$, $u=0$, which is a
solitary vortex.

\begin{figure}[!t]
\center{
 \includegraphics[width=0.55\textwidth]%
 {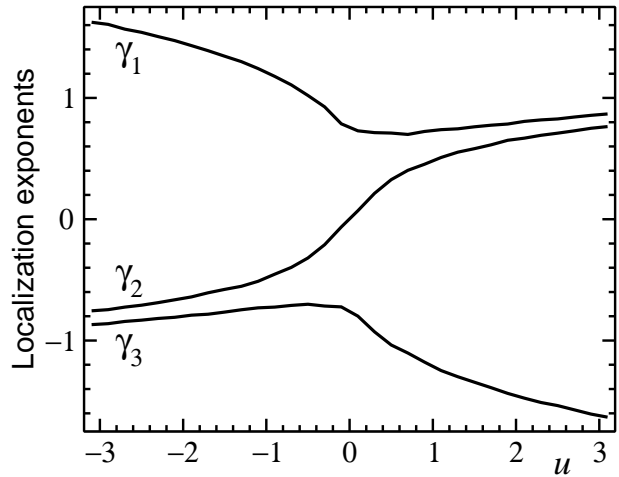}
}
  \caption{
Spectrum of localization exponents of time-independent patterns
for $\eps=1$, $q_0=-1$ derived from
equation~\eref{eq1-04}~\cite{Goldobin-Shklyaeva-PRE}.}
  \label{fig2}
\end{figure}

For $u\ne0$ the solitary patterns are still exponentially
localized [\fref{fig1}(a), $u=0.3$]. However, their localization
properties change drastically because instead of the second-order
linear ODE with respect to $\psi=\theta_x$,
equation~\eref{eq1-03}, one finds a third-order equation:
\begin{equation}
-u\theta(x)-\theta_{xxx}(x)-\xi(x)\,\theta_x(x)=q_0\,\theta_x(x)\,.
\label{eq1-04}
\end{equation}
The two symmetric modes $\theta\propto\exp(\pm\gamma x)$ and
trivial solution $\theta=const$ ($\psi=0$) turn into three modes
$\theta\propto\exp(\gamma_ix)$, $\gamma_3<0<\gamma_2<\gamma_1$, of
equation~\eref{eq1-04} with $u>0$, $q_0<0$ (see sample spectrum of
$\gamma_i$ in~\fref{fig2}; cf~\cite{Goldobin-Shklyaeva-PRE} for
details). Specifically, $\gamma_1$-mode is the successor of
$+\gamma$, $\gamma_2$-mode is the one of the trivial homogeneous
mode, and $\gamma_3$-mode is that of $-\gamma$. Thus, the upstream
flank of the localized pattern is now composed by two modes
decaying in the distance from the pattern:
\[
\theta(x)\approx\Theta_1(x)\e^{\gamma_1x}+\Theta_2(x)\e^{\gamma_2x}
\quad\mbox{ or }\quad
\psi(x)\approx\gamma_1\widetilde{\Theta}_1(x)\e^{\gamma_1x}+\gamma_2\widetilde{\Theta}_2(x)\e^{\gamma_2x},
\]
where functions $\Theta_i(x)$ and $\widetilde{\Theta}_i(x)$
neither grow nor decay over large distances. The $\gamma_1$-mode,
which disappears for $u=0$, i.e.\ $\gamma_1=0$, decays slowly for
a small finite $u$, prevails over the $\gamma_2$-mode decaying
rapidly, and, thus, determines the upstream localization
properties of the pattern. The upstream localization length
$1/\gamma_1$ can become remarkably large leading to upstream
delocalization of patterns, which can be seen in \fref{fig1}.

One should keep in mind, that consideration of solitary patterns
makes sense where such patterns can be distinguished, i.e., are
sparse enough in space. This is the case of negative
$q_0\lesssim-1$. In~\fref{fig1}, for a sample realization of
$\xi(x)$, one can see that localized patterns can be discriminated
for $\eps=1$, $q_0=-1.5$ and the localization properties are very
well pronounced for $q_0=-2.5$.

Here we would like to emphasize the fact of existence of
convective currents below the instability threshold of the
disorder-free system. These currents considerably and nontrivially
affect transport of a pollutant (or other passive scalar),
especially when its molecular diffusivity is small in comparison
to the thermal one, which is quite typical in nature (for
instance, at standard conditions the molecular diffusivity of NaCl
in water is $1.1\cdot10^{-9}m^2/s$ against the heat diffusivity of
water which is $1.3\cdot10^{-7}m^2/s$). Transport of a nearly
indiffusive passive scalar, quantified by the effective (or eddy)
diffusivity coefficient, is the object of our research, as a
``substance'' which is essentially influenced by these localized
currents and, thus, provides an opportunity to observe
manifestation of disorder-induced phenomena discussed
in~\cite{Goldobin-Shklyaeva-PRE}.

In~\cite{Goldobin-Shklyaeva-2009} we studied the problem for the
case of no advection ($u=0$) and calculated (both numerically and
analytically) the enhancement of the effective diffusivity by
disorder-induced currents; this enhancement is especially strong
for low molecular diffusivity deep below the instability threshold
of the disorder-free system (see \fref{fig3}). In this paper we
address the role of an imposed advection in this problem. The
interest to advection is provoked by its dramatic influence on
localization properties, i.e., the upstream delocalization of
convective currents that is described above. We expect this
delocalization to result in a giant increase of the effective
diffusivity for a nearly indiffusive pollutant and, particularly,
in the lowering of the mean supercriticality ($q_0$) value at
which the transition from sets of localized convective currents to
an almost everywhere intense `global' flow occurs.

\section{Effective diffusivity}\label{sec2}
In this section we describe the transport of a passive pollutant
by a steady convective flow \eref{eq1-02}; ``passive'' means that
the flow is not influenced by the pollutant. The assumption of
passiveness is practically relevant because
(biologically/chemichally) significant concentrations of a
pollutant can be very small and mechanically negligible. The flux
$\vec{j}$ of the pollutant concentration $C$ is
\begin{equation}
\vec{j}=\vec{v}\,C-D\nabla C\,,
\label{eq2-01}
\end{equation}
\noindent
where the first term describes the convective transport, the
second one represents the molecular diffusion, and $D$ is the
molecular diffusivity. The establishing time-independent
distributions of the pollutant obey
\begin{equation}
\nabla\cdot\vec{j}=0\,.
\label{eq2-02}
\end{equation}
\noindent
\Eref{eq2-02}
[with account for \eref{eq1-02}] yields a distribution of $C$
which is uniform along $z$ and is determined by
\begin{equation}
\frac{\rmd C(x)}{\rmd x}=-\frac{J}{\displaystyle D+\frac{21\,\psi^2(x)}{2\,h^2D}}\,,
\label{eq2-03}
\end{equation}
\noindent
where $J$ is the constant pollutant flux along the layer. Detailed
derivation of equation \eref{eq2-03} for $u=0$ can be found
in~\cite{Goldobin-Shklyaeva-2009} where it was performed in the
spirit of the standard multiscale method (interested readers can
consult, {\it
e.g.},~\cite{Bensoussan-Lions-Papanicolaou-1978,Majda-Kramer-1999}).
Remarkably, advection velocity $u$ is not presented in the last
equation: its direct contribution to convective currents
transferring the pollutant is small in comparison to the one of
excited convective currents. Instead, it influences the heat
transfer and, consequently, excited flows, drastically changing
properties of the field $\psi(x)$. Notice that, for the other
convective systems which we mentioned in \sref{sec1}, the result
differs only in the factor ahead of $\psi^2/D$.

Thus we come to introducing the effective diffusivity  for the
system under consideration (general ideas on the effective
diffusivity in systems with irregular currents can be found, e.g.,
in~\cite{Frisch-1995,Majda-Kramer-1999}). Let us consider the
domain $x\in[0,L]$ with the imposed concentration difference
$\delta C$ at the ends. Then the establishing pollutant flux $J$
is defined by the integral [cf \eref{eq2-03}]
\[
\delta C=-J\int\limits_0^L\frac{\rmd x}{\displaystyle D+\frac{21\,\psi^2(x)}{2\,h^2D}}\,.
\]
\noindent
For a lengthy domain the specific realization of $\xi(x)$ becomes
insignificant:
\[
\delta C=-J\,L\la\left(D+\frac{21\,\psi^2(x)}{2\,h^2D}\right)^{-1}\ra
\equiv-\sigma^{-1}J\,L\,,
\]
\noindent
Hence,
\[
J=-\sigma\frac{\delta C}{L},
\]
\noindent
which means that $\sigma$ can be treated as an effective
diffusivity.

\begin{figure}[!t]
\center{
\includegraphics[width=0.70\textwidth]%
 {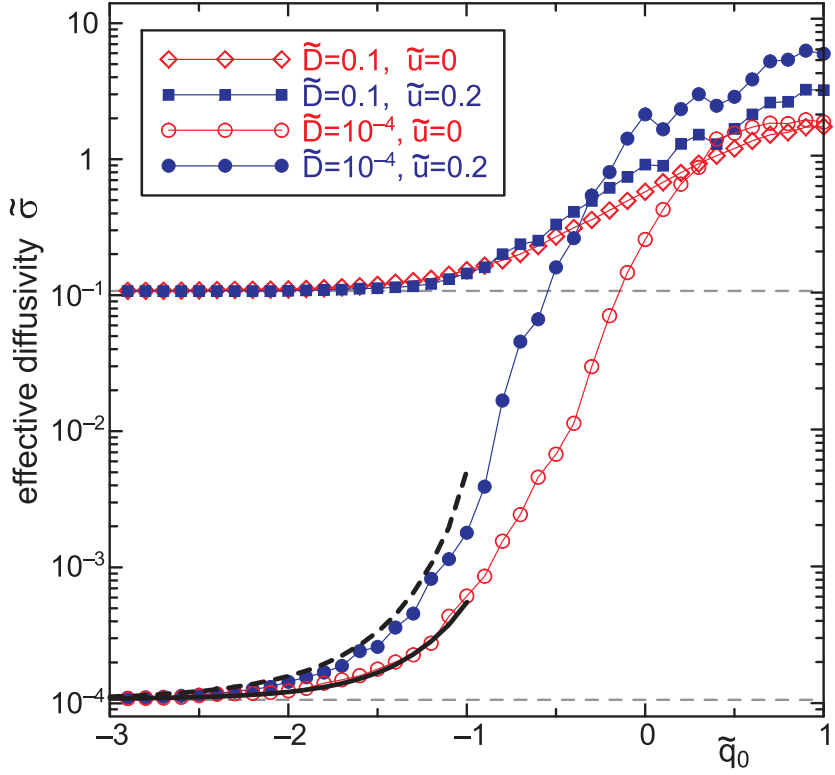}
}
  \caption{
Dependencies of effective diffusivity $\ws$ on mean
supercriticality $\wq_0$ in the presence of an imposed advection
($\wu=0.2$) and without it ($\wu=0$). The bold black solid line
represents the analytical dependence for $\wu=0$, and the dashed
line represents the one for $\wu=0.2$ (see \sref{sec3}).}
  \label{fig3}
\end{figure}

The effective diffusivity
\begin{equation}
\sigma=\la\left(D+\frac{21\,\psi^2(x)}{2\,h^2D}\right)^{-1}\ra^{-1}
\label{eq2-01}
\end{equation}
\noindent
turns into $D$ for vanishing convective flow. For small $D$ the
regions of the layer where the flow is damped, $\psi\ll1$, make
large contribution to the mean value appearing in \eref{eq2-01}
and diminish $\sigma$, thus, leading to the locking of the
spreading of the pollutant.

The disorder strength $\eps^2$ can be excluded from equations by
the appropriate rescaling of parameters and fields. As a
consequence, the results on the effective diffusivity can be
comprehensively presented in terms of $\wD$, $\ws$, $\wq_0$, and
$\wu$:
\[
\wD=\sqrt{\frac{2}{21}}\,\eps^{4/3}D,\qquad
\ws=\sqrt{\frac{2}{21}}\,\eps^{4/3}h\sigma,\qquad
\wq_0=\frac{q_0}{\eps^{4/3}}\,,\qquad
\wu=\frac{u}{\eps^2}\,.
\]
\noindent
\Fref{fig3} provides calculated dependencies of effective
diffusivity $\ws$ on $\wq_0$ for moderate and small values of
molecular diffusivity $\wD$. Concerning these dependencies the
following is worth noticing:

\noindent{\bf(a)}
For small $\wD$ a quite sharp transition of effective
diffusivity $\ws$ between moderate values and ones comparable with
$\wD$ occurs near $q_0=0$ (note logarithmic scale of the vertical
axis), suggesting the transition from an almost everywhere intense
`global' flow to a set of localized currents to take place.

\noindent{\bf(b)}
In the presence of a weak imposed advection, $\wu=0.2$, the
transition to `global' flow occurs at the value of $\wq_0$ which
is considerably lower than that without advection.

\noindent{\bf(c)}
Below the instability threshold of the disorder-free system,
where only sparse localized currents are excited, the effective
diffusion can be significantly enhanced by these currents.

\noindent{\bf(d)}
The disorder-induced enhancement of the effective diffusivity is
especially drastic in the presence of an imposed advection; e.g.,
for $\wD=10^{-4}$, $\wq_0=-1$, the effective diffusivity is
increased by one order of magnitude compared to the molecular
diffusivity without advection ($\wu=0$) and by two orders of
magnitude for $\wu=0.2$.

\section{Analytical theory}\label{sec3}
\subsection{Transport through time-independent patterns}
The effective diffusivity can be analytically evaluated for a
small molecular diffusivity ($\wD\ll1$) and sparse domains of
excitation of convective currents (the spacial density of the
excitation domains $\nu\ll1$). In~\cite{Goldobin-Shklyaeva-2009}
it was evaluated for the case of no advection,
\begin{equation}
\ws_{\wu=0}\approx\wD\left(\frac{2}{\wD}\right)^\frac{2\nu}{\gamma}\!\!,
\label{eq3-01}
\end{equation}
\noindent
where one can use the asymptotic expressions for the density of
the excitation domains $\nu$,
\begin{equation}
\nu\approx
\frac{1}{4\sqrt{1.95\,\pi}\,\eps^{2/3}|\wq_0|}
\exp\left(-\frac{1.95\,\wq_0^2}{4}\right),
\label{eq3-02}
\end{equation}
\noindent
and
\[
\gamma(\wu=0)=\eps^{-\frac{2}{3}}\left(|\wq_0|^\frac{1}{2}-\frac{1}{4}|\wq_0|^{-1}-\frac{5}{32}|\wq_0|^{-\frac{5}{2}}+\dots\right),
\]
\noindent
which are valid for $\wq_0\!<\!-1$.  The latter expression is
known from the classical theory of AL (e.g.,
see~\cite{Lifshitz-Gredeskul-Pastur-1988,Gredeskul-Kivshar-1992}).
In the following we advance the evaluation procedure realized
in~\cite{Goldobin-Shklyaeva-2009} in order to account for the
asymmetry between up- and downstream localization exponents.

Now we calculate the average
\[
\frac{1}{\ws}=\la\Bigg(\wD+\frac{\psi^2(x)}{\wD}\Bigg)^{-1}\ra.
\]
\noindent
Due to ergodicity, this average over $x$ for a given realization
of $\xi(x)$ coincides with the average over realizations of
$\xi(x)$ at a certain point $x_0$. We set the origin of the
$x$-axis at $x_0$ and find
$\ws^{-1}=\langle\big(\wD+\psi^2(0)/\wD\big)^{-1}\rangle_\xi$.

\begin{figure}[!t]
\center{
\includegraphics[width=0.68\textwidth]%
 {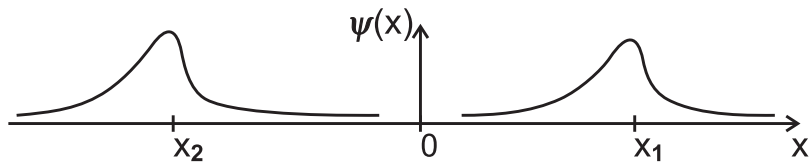}
}
  \caption{
Sketch of two localized flow patterns being nearest to the
origin.}
  \label{fig4}
\end{figure}

When the two nearest to the origin excitation domains are distant
and localized near $x_1>0$ and $x_2<0$ (see \fref{fig4}),
\begin{equation}
\psi(0)\approx\gamma_1\theta_1\e^{-\gamma_1 x_1}+\gamma_2\theta_2\e^{-\gamma_2 x_1}
  +\gamma_3\theta_3\e^{-\gamma_3x_2},
\label{eq3-03}
\end{equation}
\noindent
where $\theta_{1,2}$ and $\theta_3$ characterize the amplitude of
temperature perturbation modes excited around $x_1$ and $x_2$,
respectively. For small $\wD$ and density $\nu$, the contribution
of the excitation domains to $\ws^{-1}$ is negligible against that
of the extensive regions where flow is weak, but $|\psi|$ is still
larger than $\wD$. Therefore, one may be not very subtle with
``cores'' of excitation domains and may utilize
expression~\eref{eq3-03} even for small $x_{1,2}$:
\begin{eqnarray}
\frac{1}{\ws}=\la\frac{1}{\wD+\psi^2(0)/\wD}\ra_\xi\nonumber\\
\quad
 =\la\int\limits_0^{\infty}\rmd x_1\int\limits_0^{\infty}\rmd x_2
 \frac{p(x_1)\,p(x_2)}{\wD+\wD^{-1}(\gamma_1\theta_1\e^{-\gamma_1 x_1}
  +\gamma_2\theta_2\e^{-\gamma_2 x_1}+\gamma_3\theta_3\e^{\gamma_3x_2})^2}
 \ra_{\theta_1,\theta_2}
\label{eq3-04}
\end{eqnarray}
\noindent
where $p(x_1)$ [\,$p(x_2)$] is the density of the probability to
observe the nearest right [left] excitation domain at $+x_1$
[$-x_2$]. For probability distribution $P(x_1\!>\!x)$, one finds
 $P(x_1\!>\!x+\rmd x)=P(x_1\!>\!x)\,(1-\nu\rmd x)$,
i.e., $(\rmd/\rmd x)P(x_1\!>\!x)=-\nu P(x_1\!>\!x)$. Hence,
$P(x_1\!>\!x)=\e^{-\nu x}$, and probability density
$p(x)=|(\rmd/\rmd x)P(x_1\!>\!x)|=\nu\e^{-\nu x}$. As regards
averaging over $\theta_i$, it is important that the multiplication
of $\theta_i$ by factor $F$ is effectively equivalent to the shift
of the excitation domain by $|\gamma_{i}|^{-1}\ln{F}$, which is
insignificant for $F\sim1$ in the limit case that we consider.
Hence, one can assume $\theta_i=\pm1$ (the topological difference
between different combinations of signs of $\theta_i$ is not to be
neglected) and rewrite equation~\eref{eq3-04} as
\begin{eqnarray}
 \frac{1}{\ws}=\frac{1}{2}\int\limits_0^{\infty}\rmd x_1\int\limits_0^{\infty}\rmd x_2\,
 \nu^2\e^{-\nu(x_1+x_2)}
 \left[\frac{1}{\wD+\wD^{-1}(\gamma_1\e^{-\gamma_1 x_1}+\gamma_2\e^{-\gamma_2 x_1}
 +\gamma_3\e^{\gamma_3 x_2})^2}
 \right.\nonumber\\
 \hspace{40mm}\left.
 {}+\frac{1}{\wD+\wD^{-1}(\gamma_1\e^{-\gamma_1 x_1}+\gamma_2\e^{-\gamma_2 x_1}
 -\gamma_3\e^{\gamma_3 x_2})^2}\right].
 \nonumber
\end{eqnarray}
\noindent
For $\nu/\gamma_i\ll1$ and $\gamma_2$-mode dominating over
$\gamma_1$-mode (that is the case in \fref{fig1}), the last
formula yields
\begin{equation}
\ws\approx\wD\left(\frac{\gamma_2}{\wD}\right)^\frac{\nu}{\gamma_2}
   \left(\frac{|\gamma_3|}{\wD}\right)^\frac{\nu}{|\gamma_3|}\!\!.
\label{eq3-05}
\end{equation}
Here we assume that advection is weak and suppresses thermal
convection in a negligible fraction of the excitation centers and
the asymptotic expression \eref{eq3-02} is still valid.

\begin{figure}[!t]
\center{
\includegraphics[width=0.80\textwidth]%
 {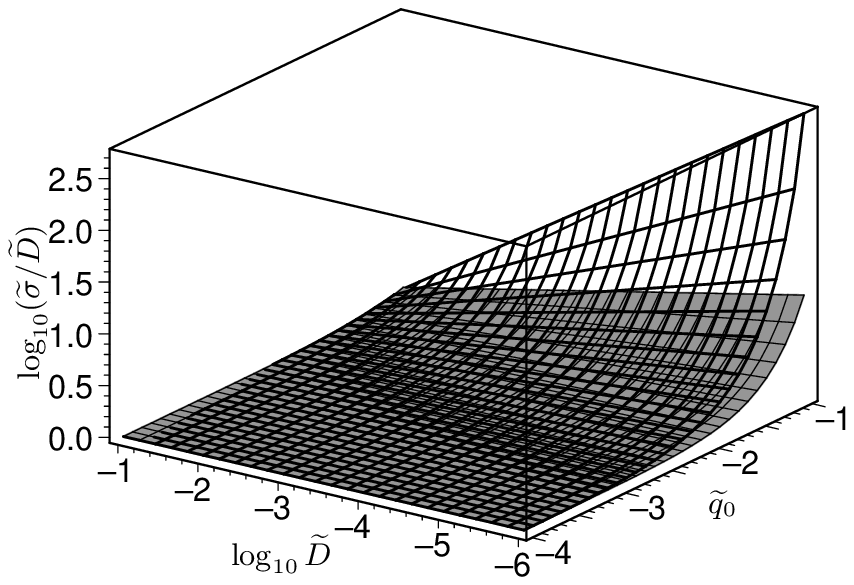}
}
  \caption{
The theoretical dependence of effective diffusivity $\ws$ on mean
supercriticality $\wq_0$ and molecular diffusivity $\wD$ in the
presence of an imposed advection, $\wu=0.2$ (black ``wireframe''
surface), and without it, $\wu=0$ (gray surface).}
  \label{fig5}
\end{figure}

Noticeable difference between equation \eref{eq3-05} for $\wu=0$,
i.e., $\gamma_2=-\gamma_3=\gamma$, and equation \eref{eq3-01} is
actually insignificant up to our approximations, because moderate
number $\gamma/2$ risen to small power $2\nu/\gamma$,
$(\gamma/2)^{2\nu/\gamma}$ which is the ratio of these equations,
approximately equals 1.\footnote{\Eref{eq3-01} is more accurate
than equation~\eref{eq3-05}, because the case of $\wu=0$ is
simpler than that we consider here and admits analytical
evaluation of integrals with a fewer number of approximations.}
For instance, in \fref{fig3}, the curves given by analytic
expressions~\eref{eq3-01} and \eref{eq3-05} with
$\gamma_2=-\gamma_3=\gamma$ are visually undistinguishable.

For small finite $\wu$, smallness of $\gamma_2$ in
equation~\eref{eq3-05} gives rise to a significant enhancement of
effective diffusivity $\ws$, which is in agreement with the
results of numerical simulation presented in \fref{fig3}.
\Fref{fig5} shows that for $\wu=0.2$ the
effective diffusivity in the presence of advection is always
stronger than without it; the larger the difference between the
effective and the molecular diffusivity the stronger advectional
enhancement of the effective diffusivity is. For instance, for
$\wD\approx 10^{-4}$, $\wq_0=-1$ the effective diffusivity in the
presence of advection $\wu=0.2$ is by factor 10 larger than
without advection, and this factor grows as $\wD$ decreases.

Noteworthy, for $\wu=0.2$ expression~\eref{eq3-05} provides
slightly overestimated value of the effective diffusivity while
for $\wu=0$ the analytical estimation is accurate. The inaccuracy
appears because in our analytical theory we ignore three factors:
(a) decrease of the spatial density of the excitation centers
owing to advectional suppression (washing-out) of weak excitation
centers; (b) for small $\wu$ the rapidly decaying upstream
$\gamma_1$-mode is significant because of the smallness of the
slowly decaying $\gamma_2$-mode; and (c) as the advection
strengthens, currents in some excitation domains disappear via a
Hopf bifurcation~\cite{Goldobin-Shklyaeva-PRE} and thus there is
non-zero probability to observe oscillatory flows for small $\wu$
even though there is no stable time-dependent solutions for
$\wu=0$. Unfortunately, inaccuracies caused by these three
assumptions can not be minimized simultaneously: the first and
third assumptions require $\wu\to0$, while the second one needs
$\wu$ to be small but finite.

\begin{figure}[!t]
\center{
\sf (a)\hspace{-5mm}
\includegraphics[width=0.464\textwidth]%
 {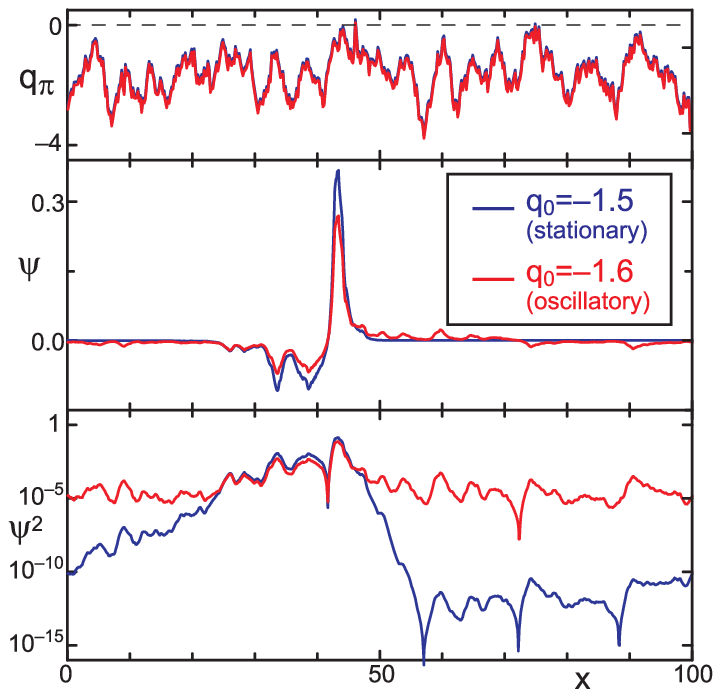}
 \quad
\sf (b)\hspace{-5mm}
\includegraphics[width=0.48\textwidth]%
 {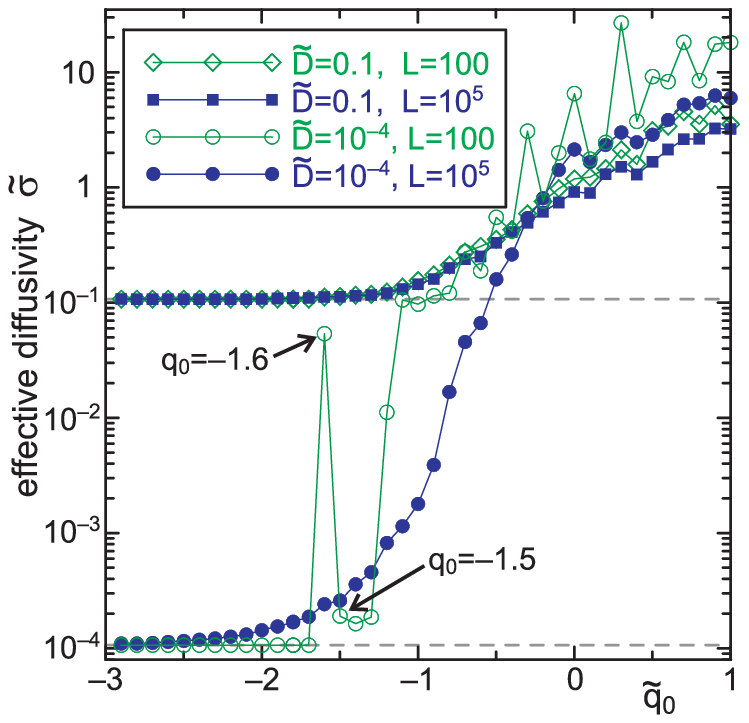}
}
  \caption{
(a) Sample of an oscillatory flow appearing for $u=0.2$, $\eps=1$:
at $\wq_0=-1.5$ the pattern is time-independent, further, as
$\wq_0$ decreases, it turns oscillatory ($\wq_0=-1.6$, shown at a
certain moment of time) and then decays for $\wq_0\lesssim-1.7$.
\\
(b) The effective diffusivity coefficients calculated over a short
domain (length $L=100$), with sample $\xi(x)$ and the patterns
plotted in graph~(a), are compared with the one in the limit of an
infinite domain ($L=10^5$).}
  \label{fig6}
\end{figure}

\subsection{Discussion of transport through oscillatory patterns}
Oscillatory localized patterns discovered in the dynamic
system~\eref{eq1-01} for non-zero $\wu$ (\fref{fig6}(a);
see~\cite{Goldobin-Shklyaeva-PRE} for details) are statistically
improbable and rare when $\wu$ is small. Notably, their relative
contribution to the effective diffusivity is much larger than
their fraction among the excited localized patterns. In
\fref{fig6}(b) one can see the soaring of the effective
diffusivity along a finite region as a localized pattern turns
oscillatory ($\wq_0=-1.6$) before disappearing
($\wq_0\lesssim-1.7$). \Fref{fig6}(a) reveals the origin of this
soaring: the oscillatory pattern is not so well localized as the
time-independent one. Indeed, the localization properties of the
oscillatory pattern of frequency $\omega$ are determined by the
following linearization of equation~\eref{eq1-01}:
\begin{equation}
i\omega\theta=-(u\theta+\theta_{xxx}+(q_0+\xi(x))\theta_x)_x\,.
\label{eq3-06}
\end{equation}
In contrast to~\eref{eq1-04}, this is already a 4th-order
differential equation, which yields four localization exponents.
The newly appeared 4th mode possesses $\gamma\propto\omega\propto
u$, i.e., decays slowly, and contributes to the downstream flank
of localized patterns (evidence of these facts is beyond the scope
of this paper and will be presented elsewhere). As well as the
$\gamma_2$-mode results in upstream delocalization of
time-independent patterns, the new mode leads to downstream
delocalization of oscillatory patterns, which appear to be weakly
localized both up- and downstream, as one sees this in
\fref{fig6}(a).

Nevertheless, owing to the smallness of the fraction of the
oscillatory patterns among all the localized patterns at small
$\wu$, their contribution to the effective diffusivity over large
domains is still negligible. This is additionally confirmed by the
accuracy of our analytical theory disregarding oscillatory
currents [equation~\eref{eq3-05}]. Meanwhile, the analytical
theory accounting for the oscillatory patterns should involve the
distribution of frequencies of excited patterns, which are to be
determined only from the nonlinear problem~\eref{eq1-01}: this is
not an analytically solvable problem.

\section{Conclusion}\label{concl}
We have studied the transport of a pollutant in a horizontal fluid
layer by spatially localized two-dimensional thermoconvective
currents appearing under frozen parametric disorder in the
presence of an imposed longitudinal advection. Though we have
considered the specific physical system, a horizontal porous layer
saturated with a fluid and confined between two nearly thermally
insulating plates, our results can be in a straightforward manner
extended to a broad variety of fluid dynamical systems (like ones
studied
in~\cite{Knobloch-1990,Shtilman-Sivashinsky-1991,Aristov-Frick-1989,Schoepf-Zimmermann-1989-1993}).
We have calculated numerically the dependence of the effective
diffusivity on the molecular one and the mean supercriticality for
a non-zero advection strength (see \fref{fig3}). The results
reveal that advectional delocalization of convective currents
greatly assists transfer of a nearly indiffusive pollutant
($D\ll1$) below the instability threshold of the disorder-free
system: the effective diffusivity can become by several orders of
magnitude larger in comparison to that without advection.

The analytical theory focusing on advectional delocalization of
localized current patterns yields results which are in a fair
agreement with the results of numerical simulation. This
correspondence confirms our treatment of importance of
disorder-induced patterns and their localization properties in
active/dissipavite media, which provoked
works~\cite{Goldobin-Shklyaeva-PRE,Goldobin-Shklyaeva-2009}.

\ack{
DSG thanks M.\,Zaks and E.\,Shklyaeva for useful discussions and
acknowledges the BRHE--program (CRDF Grant no.\,Y5--P--09--01 and
MESRF Grant no.\,2.2.2.3.8038) for financial support.}

\end{document}